\documentclass[pra,aps,twocolumn,nopacs,superscriptaddress,nofootinbib,notitlepage]{revtex4-1}

\usepackage{amsmath}  \usepackage{amssymb}  \usepackage{amsfonts}
\usepackage{mathrsfs}  \usepackage{mathtools} \usepackage{bm}  \usepackage{bbm}   \usepackage{braket}  \usepackage{color}  \usepackage{comment}  \usepackage{dcolumn}  \usepackage{enumerate}  \usepackage{epsfig}  \usepackage{gensymb}  \usepackage{graphicx}  \usepackage{indentfirst}  \usepackage{lmodern}   \usepackage{psfrag}  \usepackage{pst-all}  \usepackage{soul}    \usepackage{xcolor}
\usepackage{float}
\usepackage{cancel}
\usepackage{verbatim}
\usepackage{float} 
\usepackage[colorlinks,linkcolor=blue,citecolor=blue,urlcolor=blue,hyperindex,driverfallback=dvipdfm]{hyperref}
\usepackage[T1]{fontenc}

\def\ii{{\rm i}}  \def\ee{{\rm e}}

      \def\bb{{\bf b}}  \def\Eb{{\bf E}}      \def\fb{{\bf f}}                \def\kb{{\bf k}}      \def\Pb{{\bf P}}  \def\pb{{\bf p}}      \def\Rb{{\bf R}}  \def\rb{{\bf r}}      \def\vb{{\bf v}} 
\def\xx{\hat{\bf x}}  \def\yy{\hat{\bf y}}  \def\zz{\hat{\bf z}}    \def\rr{\hat{\bf r}}        
      
 \def\d{{\rm d}}  \def\uu{\hat{\bf u}}
\def\win{\omega_{\rm in}} \def\wout{\omega_{\rm out}} \def\kout{k_{\rm out}}

\begin{document} 
\title{Wave-mixing cathodoluminescence microscopy of low-frequency excitations}

\author{Leila~Prelat}
\affiliation{ICFO--Institut de Ciencies Fotoniques, The Barcelona Institute of Science and Technology, 08860 Castelldefels (Barcelona), Spain}
\author{Eduardo~J.~C.~Dias}
\affiliation{ICFO--Institut de Ciencies Fotoniques, The Barcelona Institute of Science and Technology, 08860 Castelldefels (Barcelona), Spain}
\author{F.~Javier~Garc\'{\i}a~de~Abajo}
\email[E-mail: ]{javier.garciadeabajo@nanophotonics.es}
\affiliation{ICFO--Institut de Ciencies Fotoniques, The Barcelona Institute of Science and Technology, 08860 Castelldefels (Barcelona), Spain}
\affiliation{ICREA--Instituci\'o Catalana de Recerca i Estudis Avan\c{c}ats, Passeig Llu\'{\i}s Companys 23, 08010 Barcelona, Spain}

\begin{abstract}
Nonlinear optical phenomena such as parametric amplification and frequency conversion are typically driven by external optical fields. Free electrons can also act as electromagnetic sources, offering unmatched spatial precision. Combining optical and electron-induced fields via the nonlinear response of material structures therefore holds potential for revealing new physical phenomena and enabling disruptive applications. Here, we theoretically investigate wave mixing between external light and the evanescent fields of free electrons, giving rise to inelastic photon scattering mediated by the second-order nonlinear response of a specimen. Specifically, an incident photon may be blue- or red-shifted, while the passing electron correspondingly loses or gains energy. These processes are strongly enhanced when the frequency shift matches an optical resonance of the specimen. We present a general theoretical framework to quantify the photon conversion probability and demonstrate its application by revealing far-infrared vibrational fingerprints of retinal using only visible light. Beyond its fundamental interest, this phenomenon offers a practical approach for spatially mapping low-frequency excitations with nanometer resolution using visible photon energies and existing electron microscopes.
\end{abstract}

\maketitle
\date{\today}

\section{Introduction}

Upon irradiation of a material with sufficiently intense monochromatic light, a polarization is induced that contains harmonics of the incident frequency \cite{B08_3}. In addition, when the material is illuminated with polychromatic light, wave-mixed components also appear at the sum and difference of the external optical frequencies. These nonlinear phenomena can be leveraged to map microscopic spatial variations in the response of different material structures, enabling suppression of the background associated with linear scattering through spectral filtering \cite{BEH97,HPR10,BPK25}. However, the spatial resolution of these methods is fundamentally limited by diffraction to approximately half the light wavelength. One approach to overcoming this problem involves the use of localized optical fields, such as those scattered from sharp tips \cite{BBH03,RA24}, which, in addition, can enhance the near-field strength and, therefore, dramatically increase the nonlinearly scattered signal \cite{AWd14,WE21,LSS25}. Nevertheless, the introduction of external nanostructures generally perturbs the intrinsic response of the specimen.

A less conventional way of generating nonlinear signals consists in mixing external light with the broadband evanescent field accompanying a swift electron. In this work, we explore this possibility as a means to perform spectromicroscopy with a combination of high spectral and spatial resolutions. Specifically, we envision the illumination and detection of visible and near-infrared light to probe an {\it idle} free-electron field in the mid- and far-infrared spectral ranges. This phenomenon could be investigated in electron microscopes equipped with external illumination capabilities, which are becoming increasingly common in the context of ultrafast electron microscopy \cite{BFZ09,FES15,PLQ15,RB16,KES18,DGH21,MWS22,ebeamroadmap}.

Electron beams (e-beams) are routinely used to achieve a high spatial resolution in electron microscopes, as they can be focused down to sub-{\aa}ngstr\"om regions \cite{NCD04,MKM08}. These probes permit electron energy-loss spectroscopy (EELS) to be performed \cite{paper149}, rendering information on the excitation modes in a specimen with a state-of-the-art spectral resolution of a few meV \cite{KLD14,HRK20}. However, EELS requires thin samples that are transparent to electrons, and it cannot resolve far-infrared excitations because they are overshadowed by the tail of the so-called zero-loss peak. The cathodoluminescence (CL) emission originating from linear scattering of the evanescent electron field provides an alternative spectromicroscopy method that can be applied to thicker samples, combined with a high spectral resolution through the determination of the emitted light wavelength \cite{paper338}. Nevertheless, CL spectroscopy suffers from a low signal-to-noise ratio due to the relatively low photon emission probability compared to energy-loss events. In addition, the detection of low frequencies remains challenging because of the poor efficiency of current detectors.

In photon-induced near-field electron microscopy \cite{BFZ09,FES15,PLQ15,RB16,KES18,DGH21,MWS22,ebeamroadmap} (PINEM), electron spectra are recorded to reveal the absorption or emission of multiple photons as the electrons traverse the near field of an illuminated specimen. A similar effect has been used in electron energy-gain spectroscopy (EEGS) to achieve a combination of high spectral and spatial resolutions by scanning the frequency of the incident light \cite{paper114,paper325,LWH19,WDS20,paper306,HRF21,paper419,MKV24}. Additionally, spectral asymmetries in the observed electron energy sidebands associated with different numbers of exchanged photons have been claimed to provide information on the nanoscale nonlinear optical response of the specimen \cite{paper347}. These methods rely on the ability to identify spectral features in the transmitted electrons, which lie beyond the capabilities of current electron spectroscopy for far-infrared excitations.

Here, we introduce wave-mixing cathodoluminescence (WMCL) as a disruptive approach for performing spectromicroscopy of low-frequency excitations, based on the nonlinear interaction between external light and the evanescent field of free electrons. Upon illumination with monochromatic light, WMCL arises from the nonlinear response of a specimen, producing outgoing light components with frequency shifts corresponding to energy losses and gains experienced by the electrons. When these frequency shifts coincide with resonances in the specimen, the electron field is enhanced and imprints distinct spectral features onto the nonlinearly scattered light. Considering a silver nanorod coated with retinal, we theoretically demonstrate that this method can reveal spectral information on the far-infrared vibrational fingerprints of this molecule through optical illumination and detection in the visible range. We further present a tutorial description of this unconventional wave mixing process for small, non-centrosymmetric particles exhibiting a nonlinear bulk response. Beyond its fundamental interest as a previously unexplored phenomenon, WMCL provides a unique opportunity to map low-frequency modes by avoiding the low efficiency of optical detectors in such a spectral range and only relying on visible-light illumination and detection.

\section{Results and Discussion}

\subsection{Theoretical framework}

\begin{figure}[htbp]
\centering
\includegraphics[width=1.0\linewidth]{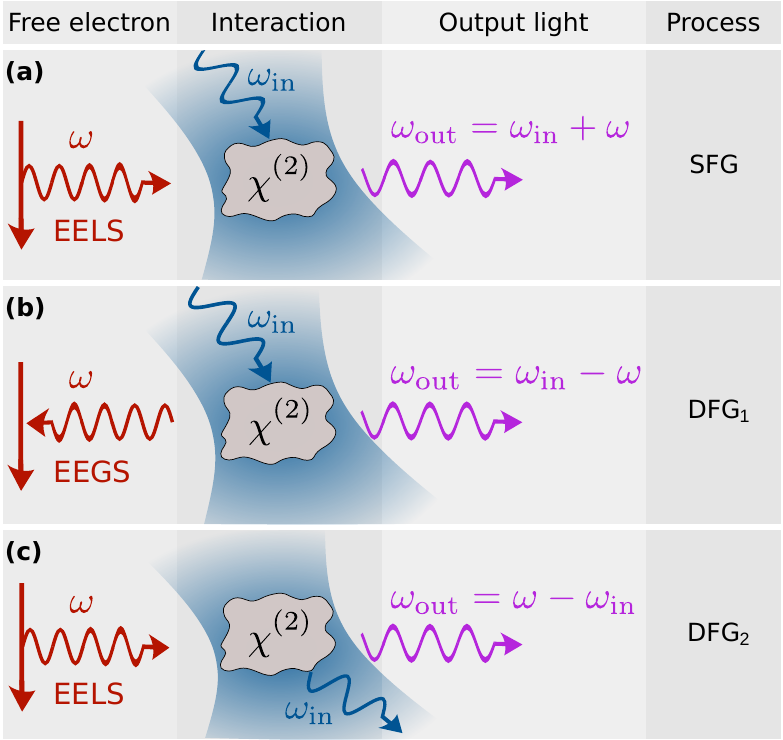}
\caption{\textbf{Wave mixing cathodoluminescence (WMCL).} We consider a specimen simultaneously exposed to light and free-electron fields. The schemes illustrate different WMCL processes involving incident photons of frequency $\win$ and the electromagnetic field component of frequency $\omega$ produced by a passing electron. The interaction is mediated by the nonlinear second-order susceptibility $\chi^{(2)}$ of a specimen, giving rise to inelastically scattered photons with an output frequency $\wout>0$. We identify three different WMCL channels: \textbf{(a)}~sum-frequency generation (SFG), with $\wout=\win+\omega$; and \textbf{(b,c)}~difference-frequency generation, with either (b)~$\wout=\win-\omega$ (DFG$_1$) or (c)~$\wout=\omega-\win$ (DFG$_2$) for $\win>\omega$ or $\win<\omega$, respectively.}
\label{Fig1}
\end{figure}

We consider a nanostructure simultaneously exposed to external monochromatic light of frequency $\win$ and a focused e-beam passing through or near the material, as depicted in Fig.~\ref{Fig1}. Wave mixing between light and the evanescent electron field can take place, assisted by the nonlinear response of the specimen and giving rise to WMCL. This phenomenon manifests in the emission of photons with an output frequency $\wout$ resulting from a linear combination of $\win$ and the different frequency components $\omega$ of the electron field. WMCL at the lowest (second) order can be anticipated to comprise three different possible channels, as summarized in Fig.~\ref{Fig1}: 
\begin{enumerate}[{(}i{)}]
\item sum-frequency generation (SFG): $\wout=\win+\omega$ (Fig.~\ref{Fig1}a);
\item difference-frequency generation at low electron frequencies (DFG$_1$): $\wout=\win-\omega$ for $\omega<\win$ (Fig.~\ref{Fig1}b);
\item difference-frequency generation at high electron frequencies (DFG$_2$): $\wout=\omega-\win$ for $\omega>\win$ (Fig.~\ref{Fig1}c).
\end{enumerate}
To map far-infrared modes (i.e., low $\omega$'s) using visible light, only channels (i) and (ii) need to be considered, but we retain channel (iii) in our analysis for completeness.

We take the external light as a plane wave of wave vector $\kb_{\rm in}$, with $k_{\rm in}=\win/c$, and an electric field distribution $\Eb^{\rm ext}_{\rm light}\,\ee^{\ii(\kb_{\rm in} \cdot \rb-\win t)}+{\rm c.c.}$, where $\Eb^{\rm ext}_{\rm light}$ is the incident field amplitude. Upon interaction with the structure, the linear optical field can be written as $\Eb_{\rm light}(\rb)\,\ee^{-\ii\win t}+{\rm c.c.}$, exhibiting a more involved spatial dependence of the total field amplitude $\Eb_{\rm light}(\rb)$.

Without loss of generality, electrons are considered to move with velocity $\vb=v\,\zz$ within an e-beam focused at a transverse position $\Rb_0$ in the $x-y$ plane. The external electric field associated with an electron has a broadband distribution given by $\Eb^{\rm ext}_{\rm electron}(\rb,t)=(2\pi)^{-1}\int d\omega\;\Eb^{\rm ext}_{\rm electron}(\rb,\omega)\,\ee^{-\ii\omega t}$ in terms of the frequency-resolved components \cite{paper149} $\Eb^{\rm ext}_{\rm electron}(\rb,\omega) = (2\ii e\omega/v^2\gamma^2)\,\ee^{\ii\omega z/v}\,\big[K_0(\zeta)\zz+\ii\gamma K_1(\zeta)\hat{\bb}\big]$, where $\gamma =1/\sqrt{1-v^2/c^2}$ is the Lorentz factor, $K_0$ and $K_1$ are modified Bessel functions evaluated at $\zeta=\omega b/v\gamma$ with $\bb=\Rb-\Rb_0$, and we introduce the notation $\rb = (\Rb, z)$ with $\Rb=(x,y)$. We disregard electron recoil (i.e., $\vb$ is assumed to be constant) as a safe approximation for the energetic e-beams under consideration. The self-consistent linear electron field $\Eb_{\rm electron}(\rb)$ generally has a more complex spatial dependence once the interaction with the specimen is accounted for.

The material responds with a nonlinear polarization density $\Pb_{\rm out}(\rb)$ at different output frequencies $\wout$ according to \cite{B08_3}
\begin{widetext}
\begin{align} \label{Pout1} 
\Pb_{\rm out}^\nu(\rb,\wout)=
\left\{
\begin{matrix}
&\chi^{(2)}(\rb,\win,\omega,\wout):\Eb_{\rm light}(\rb) \;\Eb_{\rm electron}(\rb,\omega), & \quad\quad \nu=\text{SFG}   \\
&\chi^{(2)}(\rb,\win,-\omega,\wout):\Eb_{\rm light}(\rb) \;\Eb^*_{\rm electron}(\rb,\omega)\;\Theta(\win-\omega), & \quad\quad \nu=\text{DFG}_1 \\
&\chi^{(2)}(\rb,-\win,\omega,\wout):\Eb^*_{\rm light}(\rb) \;\Eb_{\rm electron}(\rb,\omega)\;\Theta(\omega-\win), & \quad\quad \nu=\text{DFG}_2
\end{matrix}
\right. 
\end{align}
\end{widetext}
where $\chi^{(2)}$ is the position- and frequency-dependent second-order susceptibility tensor corresponding to the three different WMCL channels under consideration, indexed by $\nu$. The WMCL field amplitude can be written as $\Eb_{\rm out}^\nu(\rb,t)=(2\pi)^{-1}\int d\wout\;\Eb_{\rm out}^\nu(\rb,\wout)\,\ee^{-\ii\wout t}$, where the integral over $\wout$ indicates that it inherits a broadband character from the field of the electron (i.e., from the integral over $\omega$). The frequency-resolved emission amplitude must then incorporate the linear response of the structure at each output frequency $\wout$, which is captured by the electromagnetic Green tensor $\mathcal{G}(\rb,\rb',\wout)$, defined by \cite{NH06} $\nabla\times\nabla\times\mathcal{G}(\rb,\rb',\wout)-\kout^2\mathcal{G}(\rb,\rb',\wout)=4\pi\kout^2\delta(\rb-\rb')$, where $\kout=\wout/c$. More precisely, $\Eb_{\rm out}^\nu(\rb,\wout)=\int_V d^3\rb'\,\mathcal{G}(\rb,\rb',\wout)\cdot\Pb_{\rm out}^\nu(\rb',\wout)$, where we introduce the nonlinear polarization density defined in Eq.~(\ref{Pout1}). In the far field, the output electric field becomes $\Eb_{\rm out}^\nu(\rb,\wout)\approx\fb_{\rm out}^\nu(\Omega_{\rr})\;\ee^{\ii k_{\rm out} r}/r$, where $\Omega_{\rr}$ denotes the emission direction $\rr$ and
\begin{align} \label{fout} 
\fb_{\rm out}^\nu(\Omega_{\rr})=\int_V d^3\rb'\, g(\Omega_{\rr},\rb',\wout)\cdot\Pb_{\rm out}^\nu(\rb',\wout)
\end{align}
is the far-field WMCL amplitude. In these expressions, the $\rb'$ integral extends over the volume $V$ occupied by the nonlinear material, and the kernel in Eq.~(\ref{fout}) is implicitly defined by the far-field limit ($\kout r\gg1$) of the electromagnetic Green tensor $\mathcal{G}(\rb,\rb',\wout)\approx g(\Omega_{\rr},\rb',\wout)\;\ee^{\ii k_{\rm out} r}/r$. For a given emission direction, we use the reciprocity principle to conveniently obtain $g(\Omega_{\rr},\rb',\wout)$ from the near field produced by an incident light plane wave (see Methods).

We calculate the nonlinearly emitted energy by integrating the far-field Poynting vector over time and emission directions $\Omega_{\rr}$. Plancharel's theorem transforms the time integral into a frequency integral, which allows us to separate the emitted energy into frequency components. After dividing each of those component by the photon energy, we finally obtain the total emission probability as \cite{paper149} $\Gamma_{\rm out}^\nu=\int_0^\infty\,d\wout\,\Gamma_{\rm out}^\nu(\wout)$, where 
\begin{align} \label{Gammaout} 
\Gamma_{\rm out}^\nu(\wout)=\frac{c}{4\pi^2\hbar\wout}\int\d\Omega_{\rr}\,|\fb_{\rm out}^\nu(\Omega_{\rr})|^2
\end{align}
is the probability distribution for emitting WMCL photons normalized per electron and unit of photon frequency $\wout$ for each of the three channels $\nu=$SFG, DFG$_1$, and DFG$_2$ under consideration [see Eq.~(\ref{Pout1}) and Fig.~\ref{Fig1}].

\begin{figure*}[htbp]
\centering
\includegraphics[width=0.8\linewidth]{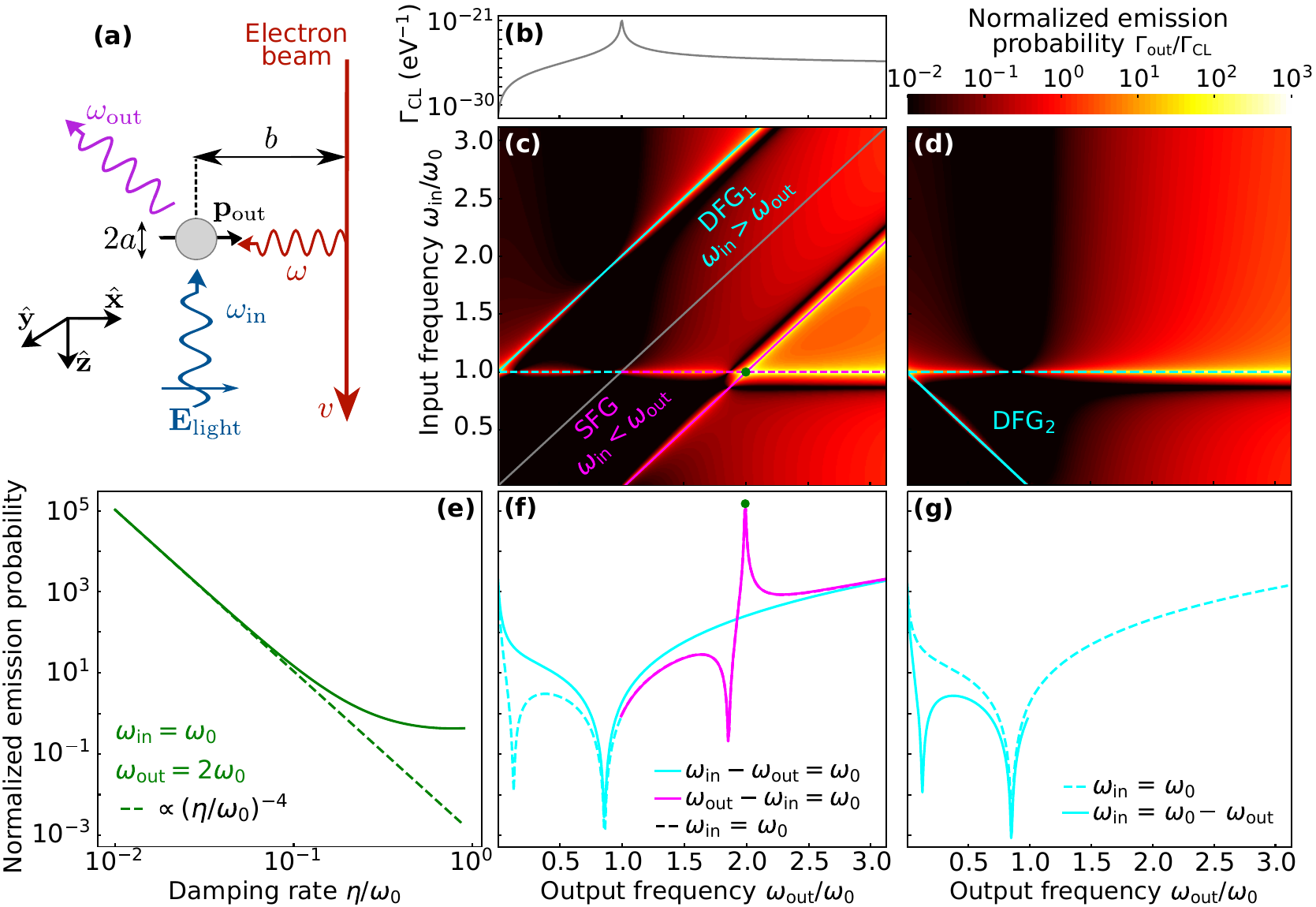}
\caption{\textbf{WMCL by a small particle.}
\textbf{(a)}~Sketch of the geometry under consideration, including a spherical particle of radius $a$ that features a dipolar resonance at frequency $\omega_0$. The electron moves with velocity $v$ and passes at a distance $b$ from the particle center. The directions of electron and light propagation are indicated by arrows relative to the lower-left frame.
\textbf{(b)}~Cathodoluminescence (CL) photon emission probability ${\Gamma}_{\rm CL}$ as a function of frequency $\wout=\omega$ normalized to $\omega_0$.
\textbf{(c)}~Probability of nonlinear WMCL photon emission $\Gamma_{\rm out}^\nu$ normalized to ${\Gamma}_{\rm CL}$ as a function of incident and output frequencies $\win$ and $\wout$ under the same conditions as in (b). The solid grey line divides regions of $\nu=$SFG ($\wout=\win+\omega$) and $\nu=$DFG$_1$ ($\wout=\win-\omega$) emission (see labels). Pink and blue lines mark the main emission features.
\textbf{(d)}~Same as (c), but for the $\nu=$DFG$_2$ process ($\wout=\omega-\win$).
\textbf{(e)}~Dependence of $\Gamma_{\rm out}^{\rm SFG}/\Gamma_{\rm CL}$ on damping $\gamma$ for $\win$ and $\wout$ at the green dot in (c) (solid curve), compared with a $\propto\gamma^{-4}$ dependence (dashed line).
\textbf{(f,g)}~Cuts of panels (c,d) along the color-coordinated lines (see legends). We set $\chi^{(2)}=10^{-10}$~m/V, $E_{\rm light}^{\rm ext}=10^8$~V/m, $a=10$~nm, $b=12$~nm, $v=c/10$, $f=1$, $\omega_0=0.1$~eV, and $\eta/\omega_0=0.01$ in all panels [see Eqs.~(\ref{epsilonLD})-(\ref{GammaCL})].}
\label{Fig2}
\end{figure*}

\subsection{Light--electron wave mixing in a small particle}

As a tutorial configuration, we study a specimen consisting of a spherical particle of small radius $a$ made of a homogeneous, non-centrosymmetric material, such that it exhibits a bulk second-order nonlinearity. For simplicity, we only consider the dipolar response for a field orientation and geometry as depicted in Fig.~\ref{Fig2}a, with a nonlinear polarization density $\Pb_{\rm out}=P_{\rm out}\xx$ associated with the $\chi^{(2)}_{xxx}$ component of the second-order susceptibility tensor.

We adopt the electrostatic limit (assuming that $a$ is small compared with the relevant optical wavelengths) and describe the material through a generic Drude-Lorentz permittivity
\begin{align} \label{epsilonLD} 
\epsilon(\omega) = 1+f\omega^2_{\rm r}/[\omega^2_{\rm r}-\omega(\omega+\ii\eta)],
\end{align}
where $f$ is a dimensionless factor, $\omega_{\rm r}$ is an intrinsic resonance frequency, and $\eta$ is an inelastic damping rate. The optical and electron fields inside the particle are then uniform and approximately given by the external fields at the sphere center times a factor $3/(\epsilon+2)$ evaluated at the frequencies $\win$ and $\omega$, respectively. The near field is thus enhanced under the condition $\epsilon=-2$, or equivalently, when the frequency approaches the sphere resonance $\omega_0 = \omega_{\rm r}\sqrt{1+f/3}$. Likewise, the kernel of Eq.~(\ref{fout}) reduces to $g(\Omega_{\rr},\rb',\wout)=\big(3k_{\rm out}^2/[\epsilon(\wout)+2]\big)\,(1-\rr\otimes\rr)$ in the electrostatic limit (see Methods). Inserting these elements in Eqs.~(\ref{fout}) and (\ref{Gammaout}), the WMCL emission probability directly becomes
\begin{align} \nonumber 
\Gamma_{\rm out}^\nu(\wout)&=3888\,\alpha\,V^2\,I^{\rm ext}\;\big|\chi^{(2)}\big|^2\;\frac{\omega_{\rm out}^3 \omega^2}{c^3v^4\gamma^2} \;K_1^2\Big(\frac{\omega b}{v\gamma}\Big) \\
&\times\frac{1}{\big|[\epsilon(\win)+2][\epsilon(\wout)+2][\epsilon(\omega)+2]\big|^2}, \label{Gammaout1}
\end{align}
where $I^{\rm ext}=(c/2\pi)|E^{\rm ext}_{\rm light}|^2$ is the incident light intensity, $V=4\pi a^3/3$ is the particle volume, $\alpha=e^2/\hbar c\approx1/137$ is the fine structure constant, and $b$ is the e-beam distance to the particle center (see Fig.~\ref{Fig2}a). A nonlinear emission channel $\nu$ is selected through the choice of $\wout$ and the frequency parameters of $\chi^{(2)}$ [see Eq.~(\ref{Pout1})]. Incidentally, from this analysis, it is clear that the WMCL probability scales as $\propto\big|\chi^{(2)}\big|^2I^{\rm ext}$.

Since both regular CL and WMCL can give rise to the emission of photons with the same frequency, the ratio of their respective probabilities becomes a relevant quantity that we explore in Fig.~\ref{Fig2}. The CL probability is also given by Eqs.~(\ref{fout}) and (\ref{Gammaout}), but taking $\wout=\omega$ and writing the polarization density as $\Pb_{\rm out}^{\rm ext}(\rb',\omega)=\Eb^{\rm ext}_{\rm electron}(-b\,\xx,\omega)\,[\epsilon(\omega)-1]/4\pi$ from the linear response to the electron field. This leads to
\begin{align} \label{GammaCL} 
\Gamma_{\rm CL}(\omega)=&\frac{3\alpha\,V^2\;\omega^5}{2\pi^3c^2v^4\gamma^2} \bigg|\frac{\epsilon(\omega)-1}{\epsilon(\omega)+2}\bigg|^2 \\
&\times\bigg[K_1^2\Big(\frac{\omega b}{v\gamma}\Big)+\frac{1}{\gamma^2}K_0^2\Big(\frac{\omega b}{v\gamma}\Big)\bigg], \nonumber
\end{align}
which agrees with previous analyses \cite{paper371}.

Figure~\ref{Fig2}c shows the $\Gamma_{\rm out}^\nu(\wout)/\Gamma_{\rm CL}(\wout)$ ratio as a function of input and output light frequencies ($\omega_{\rm in}$ and ${\omega}_{\rm out}$) for the $\nu=$SFG ($\wout>\win$) and $\nu=$DFG$_1$ ($\win>\wout$) channels, assuming a particle resonance $\hbar\omega_0=0.1$~eV, such that the e-beam can produce strong frequency shifts with a large probability (see also Fig.~\ref{FigS1} for plots of $\Gamma_{\rm out}^\nu$). We can identify some lines where the WMCL emission is boosted (marked in pink/blue for SFG/DFG$_1$ channels; see also cuts along these lines in Fig.~\ref{Fig2}f). These features correspond to resonances marked by the condition $\epsilon+2\approx0$ in the denominator of Eq.~(\ref{Gammaout1}) when either $\win=\omega_0$ (horizontal line) or $\omega=\omega_0$ (oblique lines). Note that the $\wout=\omega_0$ resonance is removed in the $\Gamma_{\rm out}^\nu/\Gamma_{\rm CL}$ ratio because $\Gamma_{\rm CL}$ is also peaked at that frequency [see Eq.~(\ref{GammaCL}) and Fig.~\ref{Fig2}b]. For completeness, Fig.~\ref{Fig2}d shows $\Gamma_{{\rm DFG}_2}(\wout)/\Gamma_{\rm CL}(\wout)$ for the alternative channel where ${\omega}_{\rm out}=\omega-{\omega}_{\rm in}$, which also exhibits resonances at $\win=\omega_0$ and $\win+\wout=\omega_0$ (dashed and solid lines). Cuts along these lines in Fig.~\ref{Fig2}g reveal a $\wout<\omega_0$ behavior similar to that of DFG$_1$ in Fig.~\ref{Fig2}f. Finally, we note that the apparent increase of $\Gamma_{\rm out}/\Gamma_{\rm CL}$ with $\wout$ in Figs.~\ref{Fig2}c,d (and $\Gamma_{\rm out}$ in Fig.~\ref{FigS1}) originates from a broad high-energy resonance associated with $\wout^3\omega^2 K_1^2(\omega b/v\gamma)$ in Eq.~(\ref{Gammaout1}).

The boosts of WMCD induced by material resonances are strongly dependent on optical damping [$\eta$ in Eq.~(\ref{epsilonLD})]. For the specific form of the permittivity in Eq.~(\ref{epsilonLD}), each resonant denominator ($|\epsilon+2|^2\approx0$) in the WMCL probability [Eq.~(\ref{Pout1})] contributes a factor $\propto(\omega_0/\eta)^2$. In particular, at the double-resonance crossing signalled by the green dot in Fig.~\ref{Fig2}c (DFG$_1$ with $\win=\omega=\omega_0$), we expect a $\Gamma_{\rm out}^{{\rm DFG}_1}\propto\eta^{-4}$, as corroborated by Fig.~\ref{Fig2}e.

\subsection{Detection of far-infrared molecular fingerprints in a plasmonic nanoparticle}
\label{Sec_cavities}

We are interested in extending the WMCL analysis to a specimen hosting multiple resonances, and in particular, one at high energy supported by a metallic nanoparticle that enhances the overall WMCL signal for input and output light in the visible range, and another one driven by molecular fingerprints of an analyte at low frequencies down to the far infrared. As a practical implementation of this idea, we consider a metallic particle coated with a molecular layer of retinal. The particle is aimed at producing a strong near-field enhancement of incident light at a visible resonant frequency. In contrast, the broadband electron field extends down to low frequencies and can be enhanced by the response of the molecules. The output WMCL signal should then be boosted for $\win$ and $\wout$ close to a nanorod resonance, while its fine structure should reveal features at frequency shifts $\wout-\win$ corresponding to the molecular fingerprints.

\begin{figure*}[htbp]
\centering
\includegraphics[width=0.8\linewidth]{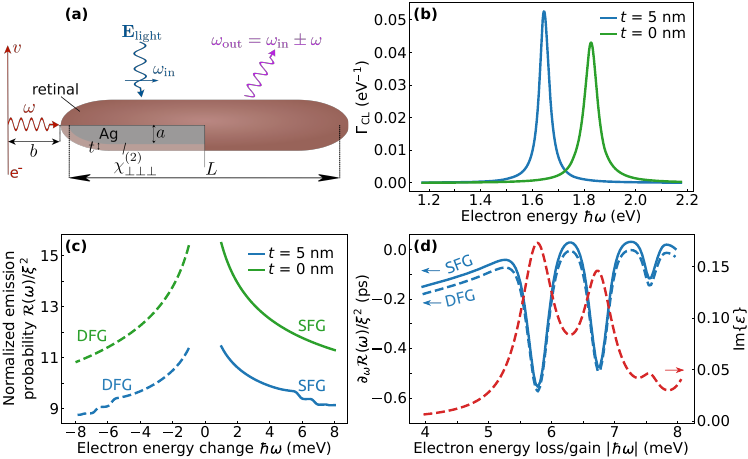}
\caption{{\bf Probing low-energy excitations through WMCL.}
\textbf{(a)}~Scheme of the structure under consideration, consisting of a silver nanorod (length $L=100$~nm, radius $a=10$~nm) coated by a retinal layer (thickness $t$). The structure is irradiated by a light plane wave of frequency $\win$ and polarization along the rod. The electron moves with a velocity $v=0.45\,c$ and passes at a distance $b=3$~nm from one end of the coated rod. SFG and DFG produce output photon energies $\wout=\win\pm\omega$ through the second-order susceptibility $\chi_{\perp\perp\perp}^{(2)}(\win,\pm\omega,\wout)$ at the silver surface.
\textbf{(b)}~CL spectra for bare ($t=0$) and coated ($t=5$~nm) silver nanorods under the geometry depicted in (a).
\textbf{(c)}~Normalized probability of SFG and DFG photon emission as a function of electron energy change. We plot the quantity $\mathcal{R}(\omega)/\xi^2$  for bare and retinal-coated nanorods, with $\mathcal{R}(\omega) = \Gamma_{\rm SFG/DFG}(\wout)/\Gamma_{\rm CL}(\wout)$, $\omega=\wout-\win$, and $\xi^2=\big|\chi^{(2)}_{\perp\perp\perp}E_{\rm light}^{\rm ext}\big|^2/A$ combining the surface area of the nanorod $A=2\pi aL$ and the incident light-field amplitude $E_{\rm light}^{\rm ext}$.
\textbf{(d)}~Comparison of $\partial_\omega\mathcal{R}(\omega)$ to the imaginary part of the permittivity of retinal as a function of absolute electron frequency change $\omega=|\wout-\win|$.}
\label{Fig3}
\end{figure*}

We take a silver nanorod (length $L=100$~nm, radius $a=10$~nm, hemispherical caps) covered by a layer of retinal (thickness $t=5$~nm), as schematically sketched in Fig.~\ref{Fig3}a. The system is illuminated by a light plane wave of frequency $\win$. An electron with velocity $v$ passes near the structure as indicated in Fig.~\ref{Fig3}a. We focus on the SFG and DFG$_1$ processes by which the electron gains or loses an amount of energy $\hbar\omega$ upon interaction with the sample, which results in the emission of a photon of frequency $\wout=\win\pm\omega$. Since silver is a centrosymmetric material, its second-order nonlinear response is only coming from the nanorod surface, and is quantified by a nonlinear surface susceptibility $\chi^{(2)}$ for which we only retain the dominant $\perp\perp\perp$ component. Symmetry breaking in the applied electric field is produced by the position of the electron, near one of the ends of the structure. In our analysis, we use the boundary-element method \cite{paper040} (BEM) to calculate the self-consistent linear fields produced by the light and the electron, from which we obtain the WMCL probability through Eqs.~(\ref{fout}) and (\ref{Gammaout}) (see Methods).

In Fig.~\ref{Fig3}b, we present the CL spectrum of the structure in Fig.~\ref{Fig3}a, exhibiting a strong dipolar resonance that shifts due to the presence of the retinal layer. For WMCL, we take the incident light frequency $\win$ to match the maximum of the resonance in this plot for a given layer thickness. In Fig.~\ref{Fig3}c, we show the probability of emitting SFG and DFG photons as a function of electron energy change, normalized to both the CL emission probability and $|\chi^{(2)}|^2$. We concentrate on the electron energy and gain region in the $-8$ to $8$~meV range, where retinal exhibits strong vibrational resonances. Such resonances are imprinted onto the WMCL spectrum of the structure with retinal, emerging as weak modulations in the normalized emission probability, which are instead absent from the spectrum for the bare nanorod. Note that the CL spectrum to which we are normalizing the WMCL signal in this energy range is dominated by the silver nanorod, so it is featureless even in the presence of a retinal layer, which only produces a nearly rigid redshift of the plasmon resonance (Fig.~\ref{Fig3}b). To make the retinal features more prominent, we take the derivatives of the WMCL spectra with respect to $\omega=|\wout-\win|$ and compare the result to the absorption spectrum of retinal (Fig.~\ref{Fig3}d), finding excellent agreement. This method can also be applied to other types of analytes. For example, for a silver nanorod coated with a thin silica layer, we find analogous features associated with vibrational resonances in the mid-infrared spectral range (see Fig.~\ref{FigS2}). We thus conclude that WMCL is capable of identifying the chemical nature of analytes characterized by low-frequency spectral fingerprints by resorting to external illumination and light detection in the visible regime.

\section{Conclusion}

In conclusion, we show that free electrons can be combined with light to produce WMCL and map optical excitations by triggering a nonlinear response associated with sum- and difference-frequency generation. Our theoretical analysis reveals that these processes can be used to reveal far-infrared spectral features in nanoscale specimens, for which no other technique is available. The successful design of experiments based on electron--photon wave mixing requires that the output signal originating from this process is larger than (and thus distinguishable from) the regular CL emission produced in the absence of external illumination. This condition depends on the second-order nonlinear susceptibility of the specimen or a neighboring structure (e.g., retinal and a silver nanorod in our illustrative example), while in addition, the WMCL intensity is proportional to the employed light intensity. The latter can be boosted by using synchronized ultrafast electron and laser pulses similar to PINEM and EEGS. Besides rendering spectral information on far-infrared excitations, WMCL should also provide quantitative measurements of the material-dependent second-order susceptibility. Although we have restricted our analysis to second-order nonlinear processes, WMCL could also be extended to higher orders, which should become dominant in small centrosymmetric samples.

\section*{Methods}
\label{Methods}

\noindent{\bf Far-field electromagnetic Green tensor from the reciprocity principle.} To calculate $g(\Omega_{\rr},\rb',\wout)$ as a function of far-field emission direction $\Omega_{\rr}$ and position $\rb'$ within a nanostructure, we consider a dipole $\pb=p\uu_1$ oscillating with frequency $\wout$, placed at $\rb'$, and oriented along the unit vector $\uu_1$. The electric field component along $\uu_2$ at a far-field position $\rb$ (with $r\gg r'$) is given by $\uu_2\cdot\mathcal{G}(\rb,\rb',\wout)\cdot\pb\approx\big[\uu_2\cdot g(\Omega_{\rr},\rb',\wout)\cdot\uu_1\big]\;p\;\ee^{\ii k_{\rm out} r}/r$. Because of reciprocity [i.e., $\uu_1\cdot\mathcal{G}(\rb,\rb',\wout)\cdot\uu_2=\uu_2\cdot\mathcal{G}(\rb',\rb,\wout)\cdot\uu_1$], this quantity must coincide with the $\uu_1$ component of the field produced at $\rb'$ by a distant dipole $p\uu_2$ placed at $\rb$. We only need to consider $\uu_2$ directions perpendicular to $\rb$ because the far field is transverse. The external field produced by such a dipole near the structure is $\kout^2p\uu_2\;\ee^{-\ii\kb_{\rm out}\cdot\rb'}\;\ee^{\ii k_{\rm out} r}/r$, where $\kb_{\rm out}=\kout\rr$. From these considerations, we formulate the following prescription to calculate the components of the tensor $g(\Omega_{\rr},\rb',\wout)$: (1) we consider a plane wave illuminating the structure with wave vector $-\kb_{\rm out}$ and unit electric field amplitude $\uu_2$; (2) we calculate the resulting self-consistent near field $\Eb(\rb')$ as a function of $\rb'$; (3) we then write $\uu_2\cdot g(\Omega_{\rr},\rb',\wout)\cdot\uu_1=\kout^2\;\uu_1\cdot\Eb(\rb')$. To obtain the full tensor, this procedure needs to be repeated for $\uu_1=\xx$, $\yy$, and $\zz$, while $\uu_2$ can be set to the p and s unit polarization vectors corresponding to the emission direction $\Omega_{\rr}$.

For a sphere in the electrostatic limit, a unit incident field $\uu_2$ produces a uniform field $3\uu_2/(\epsilon+2)$ inside the material, where $\epsilon$ is the permittivity at the frequency $\wout$ under consideration. From this identity, applying the procedure formulated above, we directly write $g(\Omega_{\rr},\rb',\wout)=\big[3\kout^2/(\epsilon+2)\big]\,(1-\rr\otimes\rr)$, where we have inserted $(1-\rr\otimes\rr)$ to account for the fact that the far field is transverse.

For the (coated) silver nanorod, we also follow the procedure above and calculate the near field upon plane-wave illumination using an implementation of the boundary-element method \cite{paper040} specialized for axially symmetric structures.

\noindent{\bf Material permittivities.} We use the tabulated permittivity of silver taken from experimental measurements \cite{JC1972}. The permittivity of retinal is modelled as a sum of six Lorentzians \cite{WFS00}:
\begin{align}
\epsilon_{\rm retinal}(\omega) = \epsilon_\infty+\sum_{j=1}^6\frac{S_j\omega_j^2}{\omega_j^2-\omega(\omega+\ii\eta_j)},
\end{align}
with parameters $\epsilon_{\infty}=2.157$, $S_1=0.018$, $\omega_1=46.6$~cm$^{-1}$, $\eta_1=5.2$~cm$^{-1}$, $S_2=0.011$, $\omega_2=54.4$~cm$^{-1}$, $\eta_2=4.7$~cm$^{-1}$, $S_3=0.001$, $\omega_3=61.0$~cm$^{-1}$, $\eta_3=2.8$~cm$^{-1}$, $S_4=0.002$, $\omega_4=66.2$~cm$^{-1}$, $\eta_4=3.4$~cm$^{-1}$, $S_5=0.001$, $\omega_5=69.3$~cm$^{-1}$, $\eta_5=2.4$~cm$^{-1}$, $S_6=0.008$, $\omega_6=90.6$~cm$^{-1}$, and $\eta_6=4.9$~cm$^{-1}$.

\section*{Acknowledgments}

This work was supported by the European Research Council (Grant No. 101141220-QUEFES), the European Commission (FET-Proactive 101017720-eBEAM), the Spanish MICINN (Severo Ochoa CEX2019-000910-S), and the Catalan CERCA Program.


%


\renewcommand{\thefigure}{S\arabic{figure}} 
\setcounter{figure}{0}

\begin{figure*}[htbp]
\centering
\includegraphics[width=0.7\linewidth]{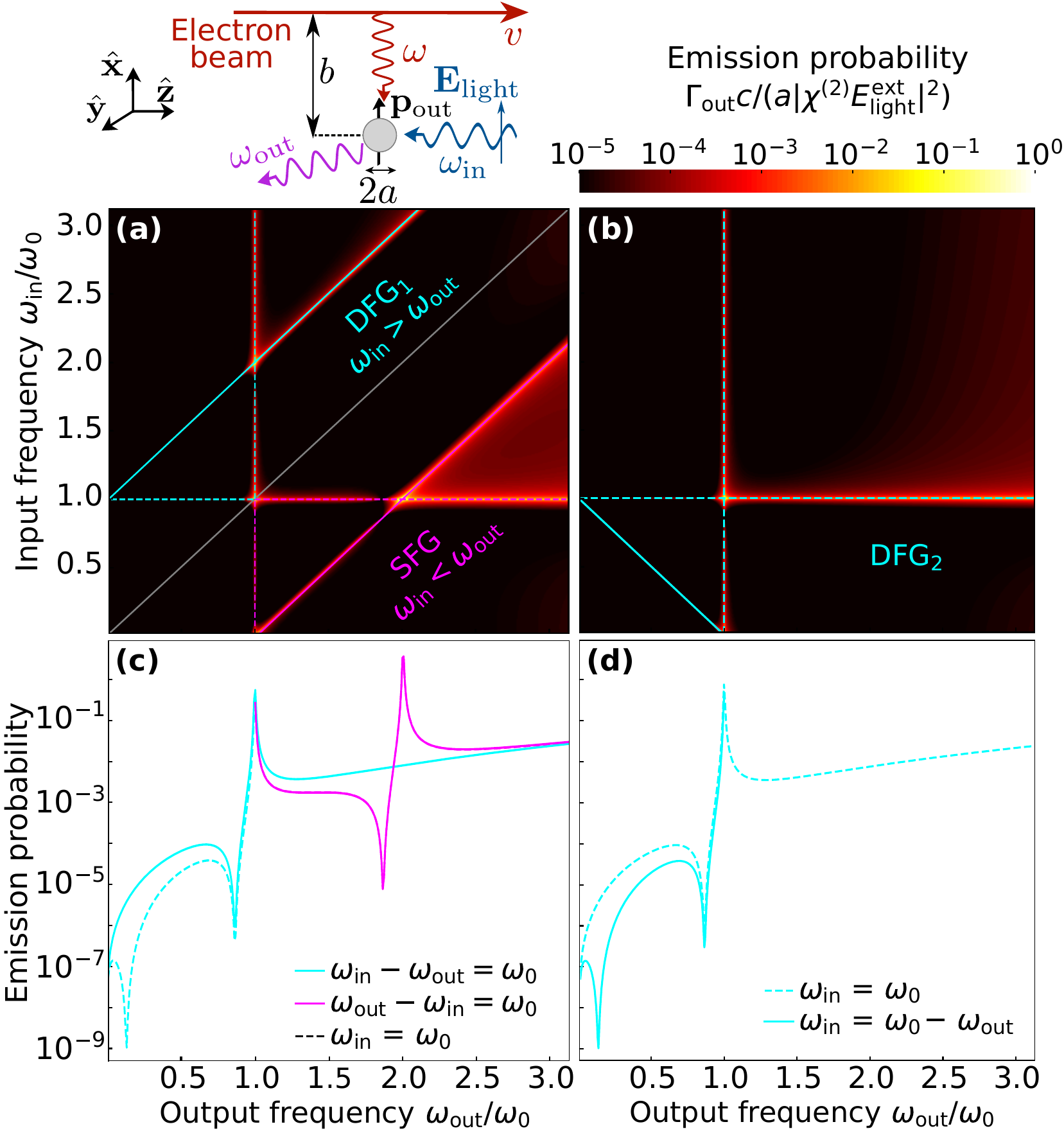}\caption{\textbf{WMCL by a small particle.} We plot the probabilities $\Gamma_{\rm out}^\nu$ in Fig.~\ref{Fig2}c,d,f,g but without normalization to $\Gamma_{\rm CL}$.
\textbf{(a)}~Probability of nonlinear WMCL photon emission $\Gamma_{\rm out}^\nu$ normalized to $c/a|\chi^{(2)}E^{\rm ext}_{\rm light}|^2$ as a function of incident and output frequencies $\win$ and $\wout$ under the same conditions as in Fig.~\ref{Fig2}b of the main text. The solid grey line divides regions of $\nu=$SFG ($\wout=\win+\omega$) and $\nu=$DFG$_1$ ($\wout=\win-\omega$) emission (see labels). Pink and blue lines mark the main emission features.
\textbf{(b)}~Same as (a), but for the $\nu=$DFG$_2$ process ($\wout=\omega-\win$).
\textbf{(c,d)}~Cuts of panels (a,b) along the color-coordinated lines (see legends).
We set $a=10$~nm, $b=12$~nm, $v=c/10$, $f=1$, $\omega_0=0.1$~eV, and $\eta/\omega_0=0.01$ in all panels.}
\label{FigS1}
\end{figure*}

\begin{figure*}[htbp]
\centering
\includegraphics[width=0.8\textwidth]{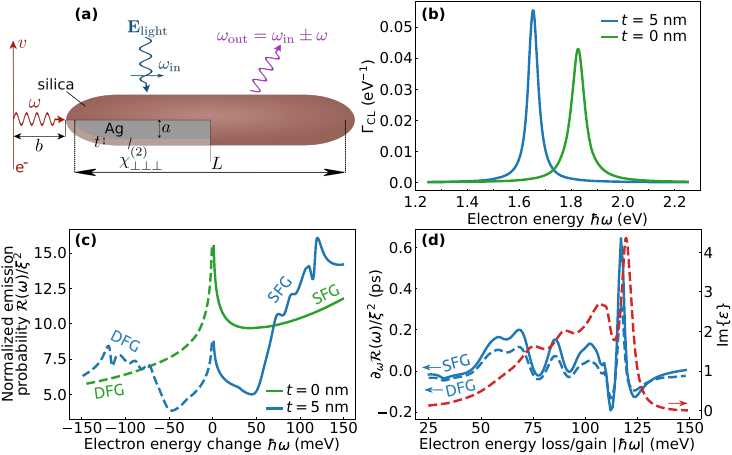}
\caption{{\bf Probing low-energy excitations in silica through WMCL.} We present data analogous to Fig.~\ref{Fig3} in the main text but with silica coating instead of retinal. The results reveal silica vibrational features in the mid-infrared range of $70-130$~meV.
\textbf{(a)}~Scheme of the structure under consideration, consisting of a silver nanorod (length $L=100$~nm, radius $a=10$~nm) coated by a silica layer (thickness $t$). The structure is irradiated by a light plane wave of frequency $\win$ and polarization along the rod. The electron moves with a velocity $v=0.45\,c$ and passes at a distance $b=3$~nm from one end of the coated rod. SFG and DFG produce output photon energies $\wout=\win\pm\omega$ through the second-order susceptibility $\chi_{\perp\perp\perp}^{(2)}(\win,\pm\omega,\wout)$ at the silver surface.
\textbf{(b)}~CL spectra for bare ($t=0$) and coated ($t=5$~nm) silver nanorods under the geometry depicted in (a).
\textbf{(c)}~Normalized probability of SFG and DFG photon emission as a function of electron energy change. We plot the quantity $\mathcal{R}(\omega)/\xi^2$  for bare and silica-coated nanorods, with $\mathcal{R}(\omega) = \Gamma_{\rm SFG/DFG}(\omega)/\Gamma_{\rm CL}(\omega)$, $\omega=\wout-\win$, and $\xi^2=\big|\chi^{(2)}_{\perp\perp\perp}E_{\rm light}^{\rm ext}\big|^2/A$ combining the surface area of the nanorod $A=2\pi aL$ and the incident light field amplitude $E_{\rm light}^{\rm ext}$.
\textbf{(d)}~Comparison of $\partial_\omega\mathcal{R}(\omega)$ to the imaginary part of the permittivity of silica as a function of absolute electron frequency change $\omega=|\wout-\win|$.}
\label{FigS2}
\end{figure*}

\end{document}